\documentclass[fleqn,twoside]{article}
\usepackage[headings]{espcrc2}

\readRCS
$Id: espcrc2.tex,v 1.2 2004/02/24 11:22:11 spepping Exp $
\usepackage{graphicx}
\usepackage[figuresright]{rotating}

\newcommand{\AmS}{{\protect\the\textfont2
  A\kern-.1667em\lower.5ex\hbox{M}\kern-.125emS}}

\title{Mini--review on Monte Carlo programs for Bhabha scattering}

\author{G.~Balossini\address[DFNT]{Dipartimento di Fisica Nucleare e Teorica,
Universit\`a di Pavia and INFN, Sezione di Pavia, \\ 
        Via A. Bassi 6, 27100 Pavia, Italy}, 
         C.~Bignamini\addressmark[DFNT], 
        C.M. Carloni Calame\address{INFN, Via E. Fermi 40, Frascati, Italy and School of Physics \& Astronomy, 
        University of Southampton,
        Southampton S017 1BJ, UK}, 
        G.~Montagna\addressmark[DFNT], 
        O. Nicrosini\address[INFN]{INFN, Sezione di Pavia, Via A. Bassi 6, 27100 Pavia, Italy}
        and
        F. Piccinini\addressmark[INFN]\thanks{The authors acknowledge support from the INTAS 
        project Nr.~05--1000008--8328
        "Higher--order effects in $e^+ e^-$ annihilation and muon anomalous magnetic moment". }}
       
\runtitle{2-column format camera-ready paper in \LaTeX}
\runauthor{G. Montagna}

\begin{document}

\begin{abstract}
We review the status of Monte Carlo 
generators presently used for simulations of the large--angle Bhabha process at electron--positron 
colliders of moderately high energy (flavour factories), operating at
centre--of--mass energies between about 1~GeV and 10~GeV. It is shown how the 
theoretical accuracy
reached by present Bhabha programs for physics at flavour factories is at the level of 0.1\% 
and, therefore, comparable with that reached about a decade ago for luminosity 
monitoring through small--angle Bhabha scattering at LEP.
\vspace{1pc}
\end{abstract}

\maketitle

\section{INTRODUCTION}

At modern electron--positron ($e^+ e^-$) accelerators built for a deeper and deeper 
understanding of the fundamental interactions between elementary particles, the machine 
(integrated) luminosity can be derived precisely by means of the relation  
$\int {\cal L} \, dt = N_{\rm obs}/\sigma_{\rm th}$, where $N_{\rm obs}$ and $\sigma_{\rm th}$ are
the number
of events and the theoretical
cross section of a given reference process, respectively. Because of the latter relation, the reference process 
must be a
reaction with clean topology, high statistics and calculable with high 
theoretical accuracy, to maintain small the total luminosity
error given by the sum in quadrature of the relative experimental and theoretical uncertainty.

For example, at high--energy accelerators LEP/SLC running in the '90s around the $Z$ pole 
to perform precision tests of the Standard Model, the process of $e^+ e^-$ production (Bhabha
scattering~\cite{Bhabha}), with the final--state leptons detected at small scattering angles, was used because 
dominated by the electromagnetic interaction and, therefore, calculable in perturbation theory, at least in principle, 
with very high theoretical accuracy. A total (experimental plus theoretical) precision of $\sim 0.05 \div 0.1\%$ was achieved at the
end of LEP/SLC operation \cite{aetal,RivNC,Lepj}, thanks to the work of different theoretical groups and 
the excellent performances of precision luminometers.

At presently running $e^+ e^-$ colliders of moderately high energy (between 
about 1 GeV and 10~GeV), such as the $\Phi$--factories DA$\Phi$NE and VEPP--2M, 
the charm--factories CESR and BEPC, as well as the $B$--factories KEK--B and PEP--II 
(globally denoted as flavour factories), the normalization reaction primarily used is the
large--angle Bhabha process \cite{lumi}. Actually, at all flavour factories, the final--state
leptons are detected at wide scattering angles, 
within a typical central acceptance region of $\sim 30^\circ \div \sim 150^\circ$, because of the absence 
of dedicated luminosity counters, for example, at small scattering angles. A precision luminosity measurement at flavour factories is of utmost importance
to perform accurate measurements of the $e^+ e^- \to {\rm hadrons}$ cross section, which is, 
in turn, a key ingredient in high--precision calculations of the running of $\alpha_{\rm QED}$ and
lepton $g-2$.
 
It is worth noting that high theoretical accuracy and, in particular,  comparison with precise
data require the development of sophisticated Monte Carlo (MC) programs, including radiative corrections
with a relative precision at the per mille level. In this review, only the large--angle Bhabha generators used at present $e^+ e^-$ colliders
will be considered. A description of the small--angle Bhabha programs available in the 
literature and used at LEP can be found in \cite{jnyr}.

\section{LARGE--ANGLE BHABHA GENERATORS} 

Since pure weak corrections are completely irrelevant in the energy range explored by 
flavour factories, the dominant part of radiative corrections, {\em i.e.} leading--log QED 
contributions, can be kept under control by means of different theoretical methods, such as 
QED Structure Functions (SF) \cite{SF} and 
Yennie--Frautschii--Suura (YFS) exponentiation~\cite{YFS}, as 
done in many applications to precision physics at LEP~\cite{RivNC}. For instance, in the generator BabaYaga v3.5 
\cite{babayaga,ips}, 
a pure QED Parton Shower (PS) is implemented, through a MC solution of the QED DGLAP 
equation satisfied by the electron SF in the non--singlet approximation. Once the electron SF is 
numerically derived, the QED corrected cross section can be simply obtained by dressing the
hard--scattering cross section of the process under consideration with a SF for any charged leg.
A remarkable advantage of this universal approach is that the four--momenta of the final--states
products (charged particles and photons) can be exclusively generated through the shower 
cascade, as in the case of the QCD PS programs for the momenta of quarks and gluons. 

\begin{figure}
\begin{center}
\includegraphics[scale=0.3]{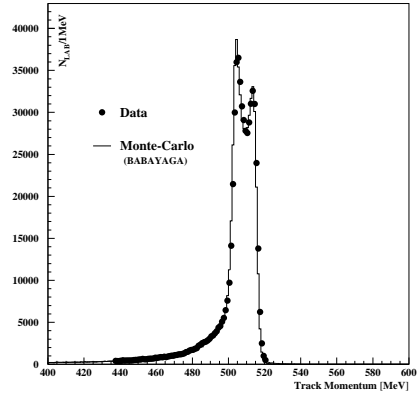}
\caption{KLOE large--angle Bhabha data for the energy distribution of the $e^+ e^-$ 
final state. From~\cite{aloisio}.}
\label{kloe}
\end{center}
\end{figure}

A comparison between BabaYaga predictions and data published by the KLOE Collaboration for
the measured energy distribution of the $e^+ e^-$ pairs produced in the large--angle Bhabha process
at DA$\Phi$NE is reported, from \cite{aloisio}, in Fig. \ref{kloe}, showing a nice agreement between
data and theory.  In spite of that, it is clear that an obvious drawback of a pure PS generator like
BabaYaga v3.5 is that it lacks the effect of  ${\cal O}(\alpha)$ non--logarithmic contributions and, therefore, the
corresponding theoretical accuracy is limited to that of a leading--log (LL) approximation. However, the physical
precision of the formulation underlying such a program can be substantially improved from about a 
per cent accuracy down to the per mille level by performing a matching of the exact 
next--to--leading (NLO) corrections with 
the multiple photon effects taken into account by the PS. This matching procedure, which is described
in detail in \cite{BabaYaga@NLO}, allows to obtain an ${\cal O}(\alpha)$ corrected cross section that 
coincides, by construction, with the exact NLO result, avoiding double counting 
and preserving resummation of leading contributions. This theoretical formulation is on the grounds
of an improved version of BabaYaga v3.5, denoted as BabaYaga@NLO, having a precision 
tag of 0.1\%. Needless to say, thanks to the matching algorithm, the theoretical error of 
BabaYaga@NLO is shifted to ${\cal O}(\alpha^2)$ 
contributions of the 
next--to--next--to--leading (NNLO) perturbative expansion. 

\begin{table*}[htb]
\caption{Relative size of different sources of correction (in per cent) to the 
large--angle Bhabha scattering 
cross section for realistic luminosity set up at $\Phi$-- and $B$--factories. From \cite{BabaYaga@NLO}.}
\label{tab2}
\newcommand{\m}{\hphantom{$-$}}
\newcommand{\cc}[1]{\multicolumn{1}{c}{#1}}
\renewcommand{\tabcolsep}{2pc} 
\renewcommand{\arraystretch}{1.2} 
\begin{tabular}{@{}lllll}

      \hline
                                      set up &     {\hskip 8pt a.}   &
      {\hskip 8pt b.}   &      {\hskip 8pt c.}   &     {\hskip 8pt d.} \\
      \hline
      $\delta_{\alpha}$                          &
      $-13.06$       &     $-17.16$       &
      $-19.10$       &     $-24.35$       \\
      \hline
      $\delta_{\alpha}^{\mathrm{non}-\log}$                  &
      $-0.39$       &     $-0.66$       &
      $-0.41$       &     $-0.70$       \\
      \hline
      $\delta_{\rm HO}$                      &
      ${\hskip 8pt 0.43}$        &     ${\hskip 8pt 0.93}$        &
      ${\hskip 8pt 0.87}$        &     ${\hskip 8pt 1.76}$        \\
      \hline
      $\delta_{\alpha^{2}L}$                    &
      ${\hskip 8pt 0.04}$        &     ${\hskip 8pt 0.09}$        &
      ${\hskip 8pt 0.06}$        &     ${\hskip 8pt 0.11}$        \\
      \hline
      $\delta_{\rm VP}$                      &
      ${\hskip 8pt 1.73}$       &     ${\hskip 8pt 2.43}$        &
      ${\hskip 8pt 4.59}$        &     ${\hskip 8pt 6.03}$        \\
      \hline
\end{tabular}\\[2pt]
\end{table*}

Theoretical approaches rather similar, at least in their basic ingredients, to that of BabaYaga@NLO 
are implemented in the MC programs MCGPJ \cite{MCGPJ} and BHWIDE \cite{BHWIDE}, both used, as BabaYaga@NLO,
at flavour factories for simulations of the Bhabha process. The former is a generator realized by a 
Dubna--Novosibirsk collaboration and includes exact ${\cal O}(\alpha)$ corrections supplemented with
higher--order LL effects taken into account through collinear SF, resulting in a theoretical precision
estimated by the authors to be better than 0.2\%. The latter is a MC code developed by the
Krakow--Knoxville group at the time of LEP operation and implements  exact ${\cal O}(\alpha)$ corrections 
matched with the resummation of soft and collinear logarithms via YFS exponentiation.
 According to \cite{BHWIDE}, the precision of BHWIDE
is conservatively estimated about 0.5\% for LEP1. However, since the theoretical ingredients of 
BHWIDE are very similar
to the formulation of both BabaYaga@NLO and MCGPJ, it is reasonable to assume that
its theoretical accuracy  for physics simulations at flavour factories is at the level of 0.1\%, 
as discussed in the following.

Concerning the theoretical precision of all the above generators, it is important to emphasize that
the bulk of the most important sub--leading ${\cal O}({\alpha^2})$ corrections, namely $\alpha^2 L$
photonic contributions enhanced by infrared logarithms, where $L = \ln (Q^2/m^2)$ is the large 
collinear logarithm, is effectively incorporated in the tools BabaYaga@NLO, MCGPJ and BHWIDE by means
of factorization of  ${\cal O}(\alpha)$ non--log terms with the leading ${\cal O}(\alpha)$ contributions
taken into account in the PS, collinear SF and YFS exponentiation approaches, as demonstrated 
in \cite{a2l}.

\section{NUMERICAL RESULTS}

To get an idea of which corrections are relevant to achieve a per mille  precision in 
simulations of the Bhabha scattering at flavour factories, we show
in Tab.~\ref{tab2} the relative effect of various contributions to the large--angle Bhabha 
cross section, when considering typical selection criteria at $\Phi$-- (set up a. and b.) and $B$--factories (set up c. and d.)~(see~\cite{BabaYaga@NLO} for details).

From Tab.~\ref{tab2}, it can be seen that ${\cal O}(\alpha)$ corrections decrease the Bhabha cross section
of about 15\% at the $\Phi$--factories and of about 20--25\% at the $B$--factories. Within the full set
of ${\cal O}(\alpha)$ corrections, non--log terms are of the order of 0.5\%, almost independently of
the centre-of-mass (c.m.) energy, as expected, and with a mild dependence 
on the angular acceptance cuts, as 
due to box/interference contributions. The effect of higher--order corrections due to multiple photon
emission is about 0.5-1\% at the $\Phi$--factories and reaches 1--2\% at the $B$--factories. The 
contribution of (approximate) ${\cal O}(\alpha^2 L)$ corrections is not exceeding the 0.1\% level, 
while the vacuum polarization increases the cross section of about 2\% around 1 GeV 
and of about 5--6\% around 10~GeV. Concerning the latter correction, the non--perturbative hadronic
contribution to the running of $\alpha$ is included in BabaYaga@NLO both in the lowest--order and
one--loop diagrams through the HADR5N routine \cite{hadr5}, that returns a data driven error, thus affecting the
accuracy of the theoretical calculation. Analogous results about the size of radiative 
corrections have been obtained recently~\cite{ggPV}
 for the process $e^+ e^- \to \gamma\gamma$, also of interest for precision luminosity studies at flavour factories.
As a whole, these results indicate that both exact ${\cal O}(\alpha)$ and higher--order corrections
(including vacuum polarization) are necessary for 0.1\% theoretical precision.

\begin{figure}[t]
\begin{center}
\includegraphics[scale=0.37]{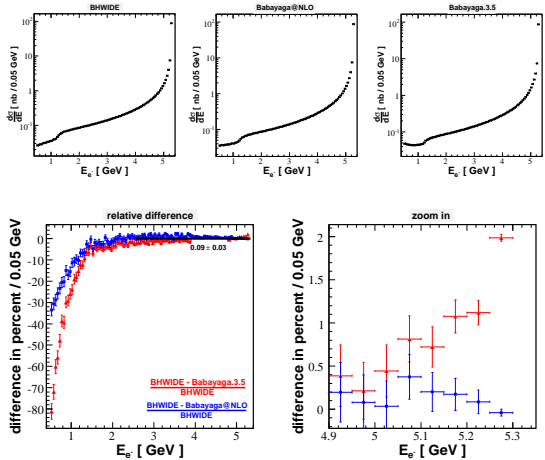}
\caption{Tuned comparisons between the results of 
BHWIDE, BabaYaga@NLO and  BabaYaga~v3.5
for the electron energy distribution in the large--angle Bhabha process at PEP--II. Courtesy of A.~Denig and A. Hafner.}
\label{andreas}
\end{center}
\end{figure}

\section{TECHNICAL AND THEORETICAL ACCURACY}

\subsection{Technical precision: tuned comparisons}

A typical procedure followed in the literature for establishing the technical precision of MC 
generators is to perform tuned comparisons between independent predictions, using the same
set of experimental cuts. An example of such tuned comparisons is given in~\cite{BabaYaga@NLO}, 
where it is shown that the agreement between the predictions of  BabaYaga@NLO
and BHWIDE at the 
$\Phi$--factories is well below 0.1\%, and that also the agreement between BabaYaga@NLO and 
LABSPV, which is a benchmark code by our group with a formulation based on collinear SF very similar to MCGPJ, is very good, below the 0.1\% level. This level of agreement, together with further 
considerations about the size of two--loop corrections discussed in the next subsection, is the reason why in
the latest publication by KLOE Collaboration about the measurement of the hadronic cross
section at DA$\Phi$NE~\cite{kloe-new} the relative uncertainty assigned to theory in the luminosity 
measurement is now 0.1\%, resulting in a total luminosity error of 0.3\%. 

Similar comparisons have been performed between the results of BabaYaga@NLO and BHWIDE by
A. Denig and A. Hafner of BABAR Collaboration, in the presence of realistic selection cuts for luminosity at PEP--II. Their studies show that the two generators agree at the 0.1\% level,  both for 
integrated and differential cross sections, in the physical and statistical significant regions.
An example of such a comparison, showing the predictions of 
BHWIDE, BabaYaga@NLO and BabaYaga v3.5  for the 
electron energy distribution (upper panels), as well as the relative differences between the results of 
BHWIDE and the two BabaYaga versions (lower panels), is given in Fig. \ref{andreas}.  

\subsection{Theoretical precision: comparisons with two--loop calculations}

In order to assess the physical precision of the generators, the 
methods typically used are {\it i)} to compare with ${\cal O}(\alpha^2)$ calculations, if the latter -- as in the
case of  Bhabha scattering -- are available in the 
literature~\cite{bdg,a2le,penin,betal,last2l,last2ll} {\it ii)} to estimate the size of unaccounted
higher--order contributions.

Concerning point {\it i)} and considering, for definiteness, the generator BabaYaga@NLO, the
strategy consists in deriving from the general formulation the ${\cal O}(\alpha^2)$ cross section, 
which can be cast in the following form
\begin{eqnarray}
\sigma^{\alpha^2} \, = \, \sigma^{\alpha^2}_{\rm SV}
+ \sigma^{\alpha^2}_{\rm SV,H} + \sigma^{\alpha^2}_{\rm HH}
\label{a2}
\end{eqnarray}
where, in principle, each of the above ${\cal O}(\alpha^2)$ contributions is affected by an uncertainty, to be properly estimated. 
In Eq.~(\ref{a2}), the first contribution is the cross section including ${\cal O}(\alpha^2)$ soft plus
virtual corrections, whose uncertainty can be evaluated by comparison with the available NNLO 
calculations. The $\sigma^{\alpha^2}_{\rm SV}$ of BabaYaga@NLO has been compared, in 
particular, with the calculation by Penin \cite{penin}, who computed the complete set of two--loop virtual
photonic corrections in the limit $Q^2 \gg m_e^2$ supplemented by real soft-photon radiation
up to non--logarithmic accuracy, and the calculations by Bonciani {\it et al.}~\cite{betal}, who
computed two--loop fermionic corrections (in the one--family approximation) with finite mass
terms and the addition of soft bremsstrahlung and real pair contributions. 

\begin{figure}
\includegraphics*[scale=0.6]{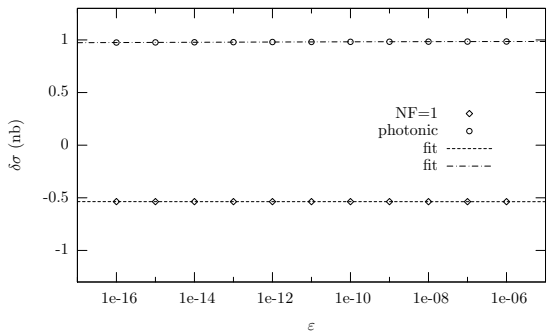}
\caption{Differences between the $\sigma^{\alpha^2}_{\rm SV}$ of BabaYaga@NLO and 
the ${\cal O}(\alpha^2)$ calculations \cite{penin} (photonic) and \cite{betal} ($N_F = 1$), 
as a function of the infrared regulator. From \cite{BabaYaga@NLO}.}
\label{2loopeps}
\end{figure}

\begin{figure}
\includegraphics*[scale=0.6]{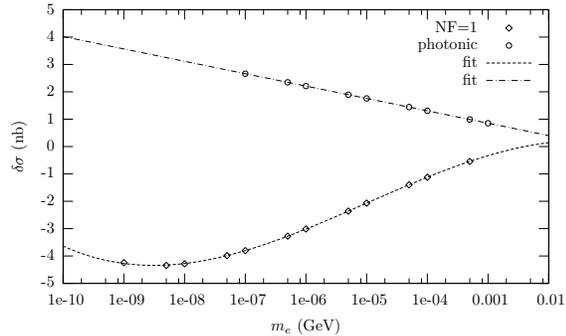}
\caption{The same as Fig. 3, as a function of a fictitious electron mass. From \cite{BabaYaga@NLO}.}
\label{2loopem}
\end{figure}

The results of such comparisons are shown in Fig.~\ref{2loopeps} and in Fig.~\ref{2loopem}, for set up a. at the $\Phi$--factories.
In  Fig.~\ref{2loopeps}, $\delta\sigma$ is the difference between $\sigma^{\alpha^2}_{\rm SV}$ of BabaYaga@NLO and the cross sections of the two ${\cal O}(\alpha^2)$ calculations, denoted
as photonic (Penin) and $N_F = 1$ (Bonciani {\it et al.}), as a function of the logarithm of the
infrared regulator $\epsilon$. It can be seen that the differences are given by flat functions,
demonstrating that such differences are infrared--safe, as expected, as a consequence of
the universality and factorization properties of the infrared divergences. In Fig.~\ref{2loopem}, 
$\delta\sigma$ is shown as a function of the logarithm of a fictitious electron mass and for a
fixed value of $\epsilon = 10^{-5}$. Since the difference with the calculation by Penin is given by
a straight line, this indicates that the two--loop soft plus virtual photonic corrections missing in BabaYaga@NLO are ${\cal O}(\alpha^2 L)$ not infrared--enhanced contributions. On the other hand, the difference with the
calculation by Bonciani {\it et al.} 
shows that the fermionic two--loop
effects missing in BabaYaga@NLO are dominated by ${\cal O}(\alpha^2 L^2)$ 
contributions. It is important to emphasize
that, as shown in detail in~\cite{BabaYaga@NLO}, the sum of the differences with the two ${\cal O}(\alpha^2)$ calculations 
does not exceed the $1.5 \times 10^{-4}$ level, for all the 
considered set up at $\Phi$-- and $B$--factories. 
The second term in Eq.~(\ref{a2}) is the cross section containing the one--loop corrections to single
hard bremsstrahlung and its uncertainty can be estimated by relying on partial results 
existing in the literature. Actually, the exact perturbative expression of $\sigma^{\alpha^2}_{\rm SV,H}$
is not available yet for full $s+t$ Bhabha scattering, but, using the results valid for 
small--angle Bhabha scattering~\cite{sabh} and large--angle $s$--channel processes \cite{sch}, the
relative uncertainty of BabaYaga@NLO in the calculation of $\sigma^{\alpha^2}_{\rm SV,H}$ can be 
safely estimated at the level of 0.05\%. The third contribution in Eq.~(\ref{a2}) is the double hard 
bremsstrahlung cross section, whose uncertainty can be evaluated by comparison with the 
exact $e^+ e^- \to e^+ e^- \gamma\gamma$ cross section. As shown in~\cite{BabaYaga@NLO}, the differences
registered between $\sigma^{\alpha^2}_{\rm HH}$ as in BabaYaga@NLO and the exact
calculation are really negligible, at the $10^{-5}$ level.

Summing all the results for the various sources of uncertainty, it turns out that the total theoretical
error in BabaYaga@NLO is $\sim 0.1$\% , when also including the uncertainty due to the
running of  $\alpha$ 
as returned by the HADR5N routine and the contribution, at a few 0.01\% 
level, of light pairs radiation~\cite{BabaYaga@NLO} , still missing in BabaYaga@NLO.

\section{CONCLUSIONS}

During the last few years, there has been a significant progress in reducing the theoretical
uncertainty in Bhabha generators used at presently running 
$e^+ e^-$ colliders down to 0.1\%. Exact 
${\cal O}(\alpha)$ and multiple photon corrections are necessary ingredients to achieve such
a precision. These corrections are implemented in three generators (BabaYaga@NLO, 
BHWIDE and MCGPJ) for the large--angle Bhabha process, 
which agree within $\sim 0.1\%$ for integrated cross sections and $\sim 1\%$ (or better)
for differential distributions~\cite{BabaYaga@NLO,MCGPJ}.

NNLO QED calculations are essential to establish the theoretical accuracy of existing generators
and, if necessary, to improve it below 0.1\%. In particular, the one--loop corrections to
single hard bremsstrahlung should be calculated for full Bhabha scattering, to get a 
better control of the theoretical precision.

For next generation $e^+ e^-$ accelerators (ILC/GigaZ), if a $10^{-4}$ accuracy is assumed, present MC Bhabha
programs need to be improved by the inclusion of weak and two--loop QED corrections, as well as beamstrahlung
and new, more precise $\Delta\alpha^{(5)}_{\rm had}$ parameterizations.

\end{document}